\documentclass[11pt]{article}
\usepackage{times}
\usepackage{sidecap}
\usepackage{geometry}
\usepackage{floatrow}
\usepackage{float}
\usepackage{hyperref}
\usepackage{graphics,graphicx}
\usepackage{subcaption}
\usepackage{epsf,wrapfig}

\usepackage{natbib,bm}
\usepackage{paralist}
\let\olditem\item
\renewenvironment{thebibliography}[1]{%
  \section*{\refname}
  \let\par\relax
  \renewcommand{\item}[1][]{\olditem[\textbullet]}%
  \inparaenum}{\endinparaenum}

\newcounter{aff}

\DeclareGraphicsExtensions{.jpg,.pdf,.png,.eps}

\usepackage{caption}
\usepackage{xcolor}

\usepackage{tcolorbox}

\bibpunct{(}{)}{;}{a}{}{,}

\usepackage{soul}

\definecolor{darkcyan}{rgb}{0.0, 0.0, 0.7}

\definecolor{boxcolor}{RGB}{193, 229, 255}
\definecolor{figboxcolor}{RGB}{193, 229, 255}
\definecolor{boxcolorgray}{RGB}{227, 227, 227}
\definecolor{darkcyan}{rgb}{0.0, 0.0, 0.7}

\sethlcolor{boxcolorgray}

\def\gtrsim{\lower 2pt \hbox{$\, \buildrel {\scriptstyle >}\over
{\scriptstyle \sim}\,$}}
\def\lesssim{\lower 2pt \hbox{$\, \buildrel {\scriptstyle <}\over
{\scriptstyle \sim}\,$}}

\def\approxlt{\lower.2em\hbox{$\buildrel < \over \sim$}}
\def\approxgt{\lower.2em\hbox{$\buildrel > \over \sim$}}

\def\be{\begin{equation}}
\def\ee{\end{equation}}
\def\bea{\begin{eqnarray*}}
\def\eea{\end{eqnarray*}}

\def\jref@jnl#1{{\rm#1}}
\def\aj{\jref@jnl{AJ}}			
\def\araa{\jref@jnl{ARA\&A}}		
\def\apj{\jref@jnl{ApJ}}			
\def\apjl{\jref@jnl{ApJ}}		
\def\apjs{\jref@jnl{ApJS}}		
\def\ao{\jref@jnl{Appl.~Opt.}}		
\def\apss{\jref@jnl{Ap\&SS}}		
\def\aap{\jref@jnl{A\&A}}		
\def\aapr{\jref@jnl{A\&A~Rev.}}		
\def\aaps{\jref@jnl{A\&AS}}		
\def\azh{\jref@jnl{AZh}}			
\def\baas{\jref@jnl{BAAS}}		
\def\jrasc{\jref@jnl{JRASC}}		
\def\memras{\jref@jnl{MmRAS}}		
\def\mnras{\jref@jnl{MNRAS}}		
\def\pra{\jref@jnl{Phys.~Rev.~A}}	
\def\prb{\jref@jnl{Phys.~Rev.~B}}	
\def\prc{\jref@jnl{Phys.~Rev.~C}}	
\def\prd{\jref@jnl{Phys.~Rev.~D}}	
\def\pre{\jref@jnl{Phys.~Rev.~E}}	
\def\prl{\jref@jnl{Phys.~Rev.~Lett.}}	
\def\pasp{\jref@jnl{PASP}}		
\def\pasj{\jref@jnl{PASJ}}		
\def\qjras{\jref@jnl{QJRAS}}		
\def\skytel{\jref@jnl{S\&T}}		
\def\solphys{\jref@jnl{Sol.~Phys.}}	
\def\sovast{\jref@jnl{Soviet~Ast.}}	
\def\ssr{\jref@jnl{Space~Sci.~Rev.}}	
\def\zap{\jref@jnl{ZAp}}			
\def\nat{\jref@jnl{Nature}}		
\def\iaucirc{\jref@jnl{IAU~Circ.}}
\def\aplett{\jref@jnl{Astrophys.~Lett.}}
\def\apspr{\jref@jnl{Astrophys.~Space~Phys.~Res.}}
\def\bain{\jref@jnl{Bull.~Astron.~Inst.~Netherlands}}
\def\fcp{\jref@jnl{Fund.~Cosmic~Phys.}}
\def\gca{\jref@jnl{Geochim.~Cosmochim.~Acta}}
\def\grl{\jref@jnl{Geophys.~Res.~Lett.}}
\def\jcp{\jref@jnl{J.~Chem.~Phys.}}	
\def\jgr{\jref@jnl{J.~Geophys.~Res.}}	
\def\jqsrt{\jref@jnl{J.~Quant.~Spec.~Radiat.~Transf.}}
\def\memsai{\jref@jnl{Mem.~Soc.~Astron.~Italiana}}
\def\nphysa{\jref@jnl{Nucl.~Phys.~A}}
\def\physrep{\jref@jnl{Phys.~Rep.}}
\def\physscr{\jref@jnl{Phys.~Scr}}
\def\planss{\jref@jnl{Planet.~Space~Sci.}}
\def\procspie{\jref@jnl{Proc.~SPIE}}

\def\farcs{\hbox{$.\!\!^{\prime\prime}$}}

\def\ion#1#2{#1$\;${\small\rm\@Roman{#2}}\relax}

\newcount\lecurrentfam
\def\LaTeX{\lecurrentfam=\the\fam \leavevmode L\raise.42ex
\hbox{$\fam\lecurrentfam\scriptstyle\kern-.3em A$}\kern-.15em\TeX}
\def\sizrpt{
(\fontname\the\font): em=\the\fontdimen6\font, ex=\the\fontdimen5\font
\typeout{
(\fontname\the\font): em=\the\fontdimen6\font, ex=\the\fontdimen5\font
}}
\definecolor{dark}{rgb}{ 0.34, 0.09, 0.49}

\usepackage{fancyhdr,url}
\usepackage{transparent}
\usepackage{tablefootnote}
\usepackage{mathpazo}

\begin{document}

\noindent
\begin{raggedright}
{\Huge\bfseries Probing cold gas and dust in quiescent galaxies across cosmic time \\[0.5em]
\Large What quenches massive galaxies in the early Universe?
}\end{raggedright}\\

\noindent
Anita Zanella$^1$, Sirio Belli$^2$, Belen Alcalde Pampliega$^3$, Manuel Aravena$^{4, 5}$ George Bendo$^6$, Rachel Bezanson$^7$, Stefano Carniani$^8$, Francoise Combes$^9$, Pierre Cox$^{10}$, Anna de Graaff$^{11}$, Daniel Espada$^{12, 13}$, Andreas L. Faisst$^{14}$, K. Decker French$^{15, 16}$, Jonathan Freundlich$^{17}$, Mahmoud Hamed$^{18}$, Zhiyuan Ji$^{19}$, Gareth C. Jones$^{20, 21}$, Vanshika Kansal$^{22, 23}$, Adarsh Kumar$^{24}$, Andrea Lapi$^{25}$, Ivanna Langan$^{26}$, Umberto Maio$^{27}$, Alessia Moretti$^{28}$, Dieu D. Nguyen$^{29}$, Mamta Pandey-Pommier$^{30}$, Gerg\"o Popping$^{31}$, Prachi Prajapati$^{32}$, Guillermo Quintana-Lacaci$^{33}$, Pierluigi Rinaldi$^{19}$, Raghvendra Sahai$^{34}$, Justin Spilker$^{35}$, Sandro Tacchella$^{20, 21}$, Sabine Thater$^{36}$ Francesco Valentino$^{37, 38}$, Livia Vallini$^1$, Jacco van Loon$^{39}$, Enrico Veraldi$^{25}$, Christina C. Williams$^{40}$

\begin{tcolorbox}[boxsep=3pt,left=0pt,right=0pt,top=1pt,bottom=1pt,colframe=boxcolor,colback=boxcolor]
\justify
\textbf{Abstract:} 
The discovery of massive quiescent galaxies at high redshift by JWST has revealed that star formation can cease rapidly in the early Universe, yet the physical mechanisms responsible for quenching remain unknown. A key missing element is the systematic measurement in statistically representative quiescent samples of the cold gas and dust content, which are crucial for distinguishing quenching mechanisms and evolutionary pathways.
Such observations are currently not feasible because state-of-the-art sub-millimeter facilities (e.g., ALMA) cannot routinely detect the faint molecular gas and dust reservoirs of quiescent galaxies, nor pinpoint outflowing molecular gas. A next-generation facility (ALMA2040) capable of providing transformative access to the cold interstellar medium of quiescent galaxies across redshift is therefore essential. In particular, it will allow the community to probe the molecular gas content of quiescent galaxies at cosmic noon and beyond by targeting the CO, [CI], and [CII] emission lines, as well as unveiling the physical conditions of the cold interstellar medium by constraining the CO spectral line energy distribution. It will also reveal the dust content of the passive population, a crucial ingredient to constrain the different quenching mechanisms. Finally, it will enable probing the molecular gas kinematics and identify molecular outflows driven by supermassive black holes. When combined with the rich legacy of JWST, ELT, and SKAO surveys, ALMA2040 will deliver a comprehensive, multi-phase view of feedback and quenching, allowing us to establish when, how, and why galaxies shut down their star formation across cosmic time.
\end{tcolorbox}

\newpage{}

\section{Scientific context and open questions}
\label{sec:context}

\begin{figure}[t!]
    \centering
    \includegraphics[width=\linewidth]{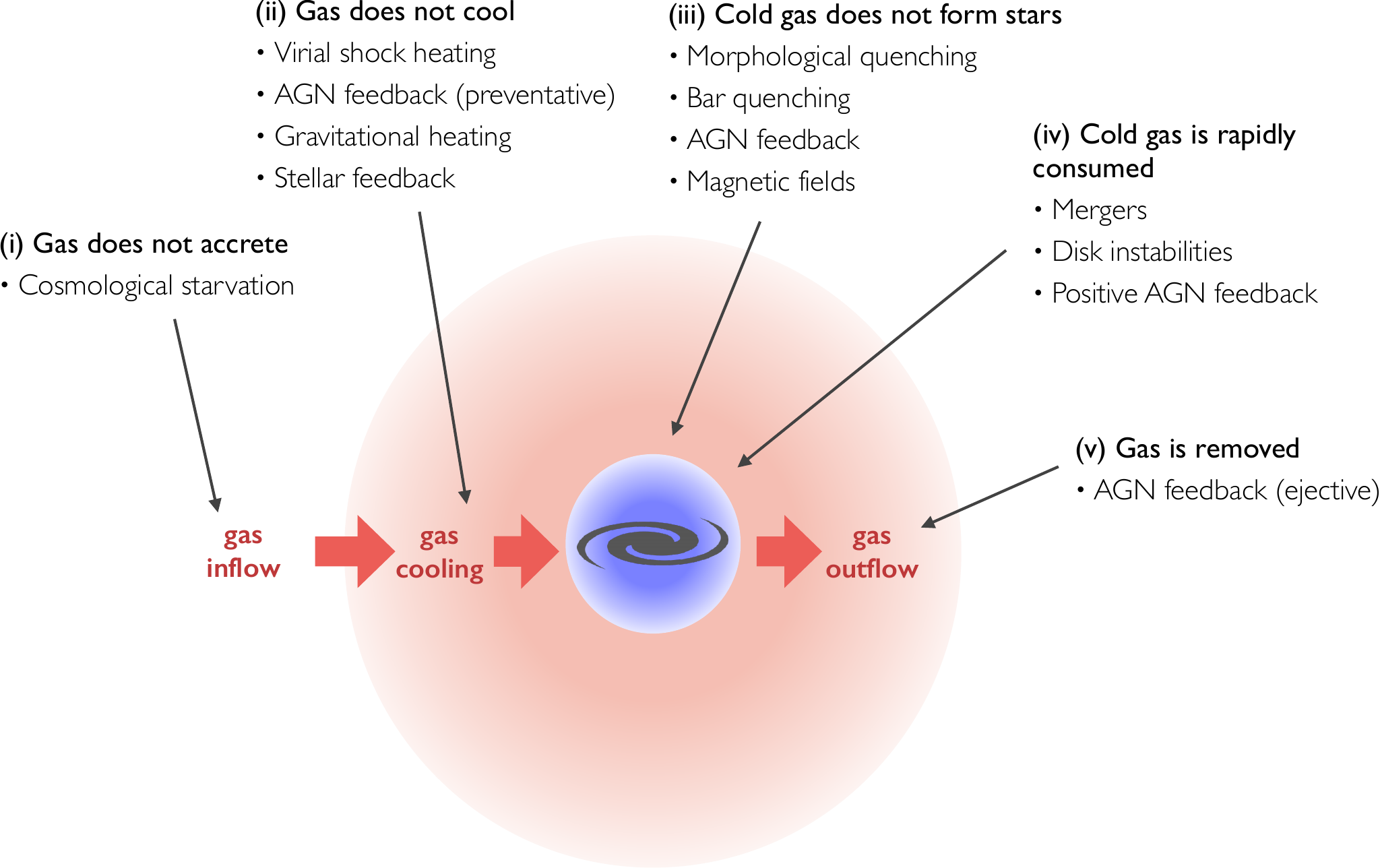}
    \caption{Schematic view of possible quenching mechanisms of massive galaxies. Various processes leave a different imprint on the molecular gas and dust reservoirs of quiescent galaxies. Figure adapted from \cite{Man2018}.}
    \label{fig:quenching_mechanisms}
\end{figure}

The transition of galaxies to a state of low star formation  (i.e., quiescence) represents a crucial open issue in the current paradigm of galaxy evolution. 
In the Lambda Cold Dark Matter cosmological scenario structures grow hierarchically, and therefore the most massive dark matter halos are the last to form. However, observations in the local universe reveal that the most massive galaxies host the oldest stellar populations, meaning that they experienced a rapid, early growth followed by a long period of quiescence.
Explaining this ``anti-hierarchical'' stellar mass assembly remains a central challenge for galaxy formation models.
In recent years, this issue has become even more pressing by the unexpected discovery, enabled by JWST, of massive quiescent galaxies at $z>3$ when the universe was only 1-2~Gyr old \citep{Carnall2023, deGraaff2025, Weibel2025}. 
The star formation activity in these systems must have been rapidly quenched despite the high cosmological gas accretion rates and large molecular gas fractions that characterize these epochs.
Understanding the physical nature of this quenching process remains a major goal of extragalactic astronomy. 
Is star formation quenched because the gas is expelled from the galaxy or because the gas properties (e.g. temperature, density, turbulence) prevent the formation of molecular clouds and new stars?
Different physical models of quenching have been developed (Figure \ref{fig:quenching_mechanisms}), acting differently on the baryon cycle: (i) reducing the gas accretion; (ii) heating the molecular gas or preventing gas cooling; (iii) retaining cold gas, but suppressing or preventing its conversion into stars; (iv) rapidly consuming the gas reservoirs; or (v) ejecting existing gas \citep{Man2018, Whitaker2026}. Isolating these physical models requires disentangling internal, secular quenching from external environmental drivers. This distinction becomes especially important in group and cluster environments, where the quenching of star formation is dictated by a complex interplay of both.

\begin{figure}[t!]
    \centering
    \includegraphics[width=\linewidth]{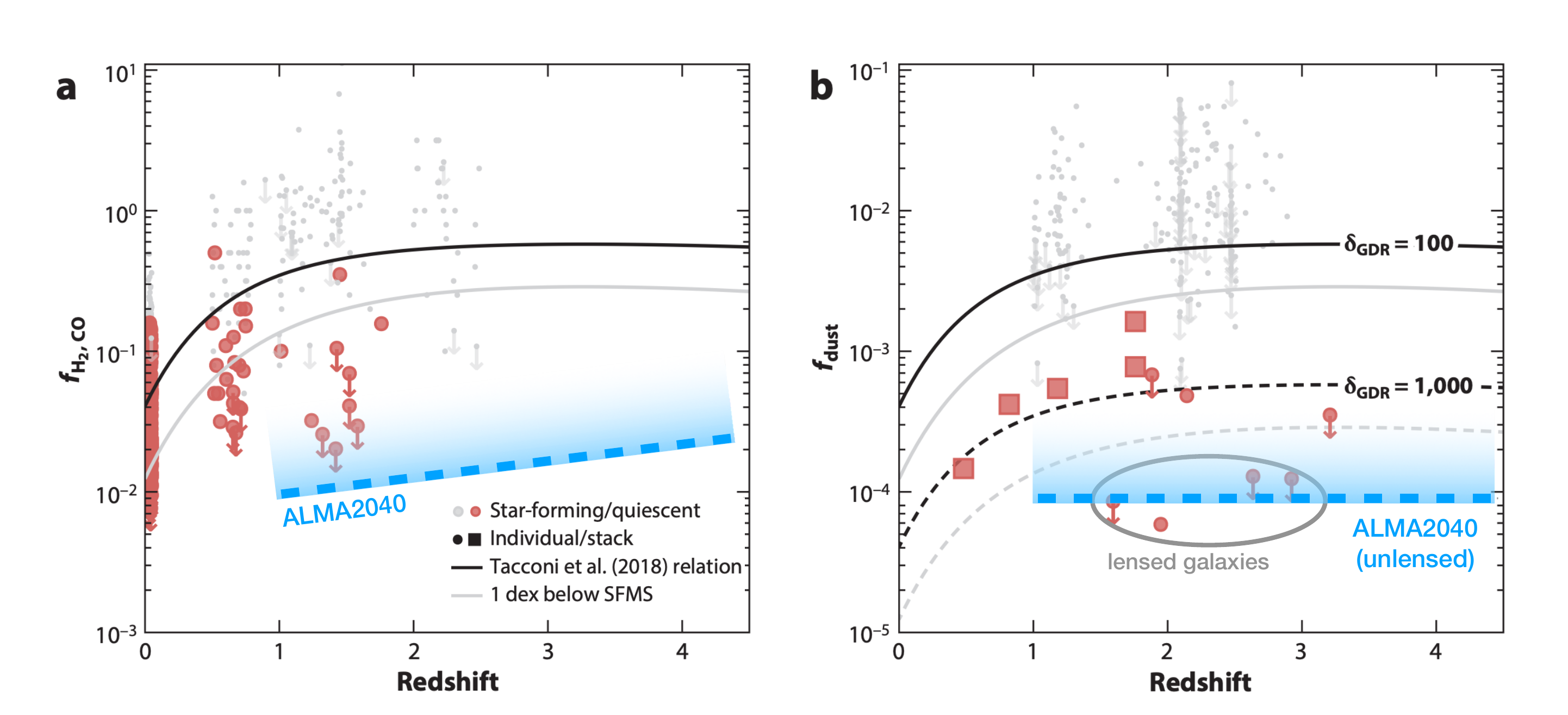}
    \caption{Over the last decade, ALMA has revealed the molecular gas reservoirs of quiescent galaxies by using low-$J$ CO lines and dust continuum tracers. \textbf{Left panel:} molecular gas fraction of quiescent galaxies estimated from CO lines across redshift. The gas fraction for galaxies on the star-forming main sequence \citep{Tacconi2018} is  shown as a black curve, and the same relation shifted 1 dex downward is shown as a gray curve. Measurements are limited to $z < 2$ and require substantial investment of ALMA time. The increased sensitivity of ALMA2040 will enable probing molecular gas fractions $\mathrm{f_{gas} \sim 2\%}$ for galaxies with stellar mass $\mathrm{M_\star \sim 10^{10} - 10^{11}}\, M_\odot$ at $z > 1$ (blue dashed line). The use of the brighter [CII] line at higher redshift produces a sensitivity that is only mildly decreasing with redshift. \textbf{Right panel}: dust fraction of quiescent galaxies across redshift. Typical gas-to-dust ratios for star-forming (black curves) and quiescent (gray curves) galaxies are indicated. Current measurements, which are limited to $z < 3$, rely mostly on stacks and require multiple assumptions about dust properties (e.g., opacity, emissivity, temperature), while individual measurements of gravitationally lensed galaxies typically yield non-detections. ALMA2040 will enable measurements for individual unlensed galaxies down to $\mathrm{log\, f_{dust} \sim -4}$ (blue dashed line). The sensitivity is nearly constant with redshift due to the negative $k$-correction, i.e., the shifting of the dust emission peak into the ALMA frequency range at higher redshifts. This figure is adapted from \cite{Whitaker2026}. }
    \label{fig:literature}
\end{figure}

Obtaining a detailed view of the gas content in quiescent galaxies, including its spatial distribution and physical conditions, is key for understanding the quenching process and providing a physically motivated basis to galaxy formation models.
Over the last decade, ALMA observations have led to enormous progress in our understanding of the gas cycle in star-forming galaxies across cosmic time, but quiescent galaxies remain elusive because of the difficulty of detecting the faint dust and CO emission from their cold gas due to the lack of sufficient sensitivity (Figure \ref{fig:literature}). Beyond the local Universe, only a few tens of quiescent galaxies have been characterized at $0.2 < z < 1$ and a handful at $1 < z < 2$ despite substantial investment of ALMA time \citep{Suess2017, Spilker2018, Williams2020, Belli2021, Bezanson2021, Whitaker2021b, Zanella2023, Setton2025}. Individual detections of CO or [CII] emission lines are almost absent at $z > 2$ \citep{Umehata2025, Valentino2026}, and even meaningful non-detections are rare at these redshifts \citep{Scholtz2024, Ji2024}. The lack of dust and gas measurements in distant quiescent galaxies is recognized as one of the critical limitations to our current understanding of the quenching process \citep{Whitaker2026}.

\subsection{ALMA2040: science case}
\label{subsec:science_case}

ALMA2040 will enable three crucial measurements in large samples of quiescent systems at $1 < z < 5$: (1) Emission line fluxes to probe molecular gas reservoirs and physical conditions, directly linked to the gas availability and ability to form stars; (2) Dust continuum emission to uncover the processes governing dust growth and destruction; (3) Gas kinematics to detect large-scale molecular outflows, if present, and to quantify AGN-driven gas removal as a function of stellar mass. These measurements will in turn allow the community to answer key scientific questions: (a) Do quiescent galaxies still retain the fuel for star formation? (b) If the molecular gas reservoir is substantial, why is it not forming stars? What are its physical conditions and its spatial distribution? (c) If the gas was removed, what are the physical processes responsible for this removal?

\subsubsection{Characterizing the molecular gas reservoirs}
\label{subsec:mol_gas}

Understanding the physical mechanisms that shut down star formation requires observations of quenched systems to characterize their cold gas reservoirs and the physical conditions of their interstellar medium (ISM). Recent measurements of low-$J$ CO transitions such as CO(2–1) and CO(3–2), which are classical tracers of the cold molecular gas, have revealed substantial gas reservoirs in quenched galaxies out to $z \sim 1.4$ \citep{French2015, Suess2017, Belli2021, Bezanson2021, Zanella2023}. Reported gas fractions ($\mathrm{f_{gas} = M_{H2}/M_\star}$) can reach up to 20\%, indicating that a sizable amount of star-forming fuel may persist after star formation has ceased. These galaxies must therefore have a low star formation efficiency for reasons that are still unclear. The low efficiency may be caused by high molecular gas temperatures and low molecular gas densities or by dynamical stabilization against collapse \citep{French2018b, vandeVoort2018, Zanella2026, Maio2026}. 

Current samples are limited in size and heterogeneous in stellar mass, stellar population age, and time since quenching ($t_\mathrm{q}$). They are typically biased towards massive systems ($\mathrm{M_\star \gtrsim 10^{11}\, M_\odot}$) and/or recently quenched ($t_\mathrm{q} \lesssim 1\, \mathrm{Gyr}$) galaxies, and little is known about their environment, preventing robust conclusions about the physical drivers of quenching \citep{Noble2019}. 
Furthermore, the current detections may represent only the gas-rich tail of the population, further biasing our view of quenching, while measurements of molecular gas in quiescent galaxies at $z > 2$ are almost absent \citep{Umehata2025}, despite the fact that this epoch is critical for witnessing quenching in action. 

The [CII] emission line has been proposed as an alternative molecular gas tracer in star forming galaxies \citep{Zanella2018, Madden2020, Zhao2024}. The [CII] line can be substantially brighter than CO lines, and is particularly advantageous at $z>3$, where the reduced contrast against the CMB suppresses the low-\textit{J} CO lines \citep{DaCunha2013}. The main limitations of this line are the poorly constrained [CII]-to-H$_2$ conversion factor, especially for quiescent galaxies, and the fact that [CII] does not trace exclusively the molecular gas, but can also trace gas in the neutral atomic and ionized phases. These issues are independent of those affecting the use of CO, making [CII] an important independent proxy of the molecular gas mass.
Observations of the [CII] line are promising for studying the cold gas reservoirs of high-redshift quiescent galaxies, as shown by a few recent studies \citep{DEugenio2023, DEugenio2026}, including a detection in a quiescent galaxy at $z = 7.3$ \citep{Valentino2026}. However, a systematic search for [CII] in quiescent systems is still missing and it is hardly affordable with current ALMA capabilities due to the long required integration times.

An additional probe of the total molecular reservoir is provided by [CI], which is particularly valuable in quenched and AGN-irradiated systems. In regions where an enhanced radiation field or cosmic-ray flux photodissociates CO, a growing fraction of the H$_2$ becomes CO-dark and is traced instead by [CI] \citep{Papadopoulos2004, Bisbas2017}. Since the dust-to-gas ratio and $\alpha_{\rm CO}$ are both expected to depart from star-forming values after quenching (Sect.~1.1.2), a [CI]-based cross-check removes a key systematic in the gas-mass and outflow-rate budgets. Cross-calibrating multiple gas tracers in the same systems will be crucial to obtain accurate calibrations and molecular gas estimates.

ALMA2040 sensitivities will enable systematic measurements of molecular gas in complete samples (in terms of stellar mass, quenching time, and redshift) of quiescent galaxies at $z \sim 1$–5 (and possibly even $z > 5$) using emission lines such as CO, [C II], and [C I].  Molecular gas fractions down to $\mathrm{f_{gas} \sim 2\%}$ in galaxies with stellar mass $\mathrm{M_\star \sim 10^{10} - 10^{11}\, M_\odot}$ will be probed (see Section \ref{sec:technical_req}). These observations will reveal how quenching happens and whether it is maintained at high redshift, thus allowing us to discriminate between permanent quiescence and rejuvenation. By complementing ALMA2040 results with observations from the UV to the near-infrared it will also be possible to measure the gas fraction and ISM state as a function of time since quenching (see Section \ref{sec:synergies}). In particular, using the star formation histories and stellar ages derived from near-infrared spectroscopy (e.g., from JWST or ELT), it will be possible to reconstruct the evolution of the gas reservoir after shutdown \citep[e.g.,][]{Scholtz2024} and pinpoint the mechanisms driving the decrease of star formation by understanding whether the gas disappears rapidly during quenching, declines slowly afterwards, or remains present but inefficient for several hundred Myr.

In addition, a higher sensitivity will make it possible to target multiple CO transitions in individual quiescent galaxies, allowing the construction of CO spectral line energy distributions (SLEDs). These SLEDs constrain the excitation state and multi-phase structure of the molecular gas and, when combined with [C I] and dust measurements, provide information on its characteristic density, temperature, and radiation environment. Furthermore, the shape of the CO SLED will be informative regarding the presence of shock heated molecular gas and/or the possible impact of low-luminosity AGN warming up the molecular gas \citep{Vallini2019, Esposito2024}.
At present, only a handful of quiescent galaxies at $z \lesssim 1$ have been observed in two or more CO transitions with ALMA \citep{Zanella2026, Donofrio2026}. Increased sensitivity will allow CO SLED measurements for substantially larger samples out to higher redshifts, enabling a direct assessment of the mechanisms that keep galaxies quenched, whether through elevated gas temperatures, low gas densities, or other suppression processes.

ALMA can, in principle, spatially resolve massive quiescent galaxies at $z \sim 1$–5, because their effective sizes are of the order of 0.5$-$5~kpc \citep{Ji2026}, corresponding to $\sim0\farcs1-0\farcs7$. However, spatially resolved studies are currently unfeasible due to sensitivity limitations.
As a consequence of the increased sensitivity of ALMA2040 it will be possible to spatially resolve the molecular gas emission and probe its morphology. This is particularly important for studying interactions and ram pressure stripping, which may play a major role in quenching galaxies in the densest environments \citep{Xu2026}. Spatially resolving the molecular gas reservoirs will also provide constraints on the molecular gas surface density, crucial for understanding deviations from star formation scaling relations. Furthermore, probing the size of different CO $J$ transitions will be a diagnostic of the molecular gas density as a function of the galactocentric radius, a key test for inside-out quenching \citep{Tacchella2015b}.

Finally, the increased sensitivity and wide bandwidth of ALMA2040 will enable the detection of emission lines (e.g., HCN, H$_2$O) from denser gas than the ISM probed by CO, [CII] or [CI]. Despite being 10 - 50 times fainter than CO, their detection at least in the brightest galaxies will open the possibility to investigate the presence and properties of the denser gas reservoirs in quiescent galaxies, a study that so far has only been conducted in the local Universe for a handful of post-starburst systems \citep{French2018}.

\subsubsection{Constraining the dust content and quenching mechanisms}
\label{subsec:dust}

Dust continuum emission in quiescent galaxies has often been used as an indirect tracer of their molecular gas reservoirs, typically by assuming gas-to-dust ratios similar to those in star-forming systems ($\delta_\mathrm{GDR} \sim 100$). However, most existing constraints rely on stacking analyses \citep{Gobat2018, Magdis2021}, as only a small number of individual quiescent galaxies at $z \sim 1$–3 have been detected in dust continuum \citep{Whitaker2021a, Lee2024}.
Recent theoretical work \citep{Whitaker2021b, Lorenzon2025a} has questioned the validity of using dust as a proxy for molecular gas in quenched systems, suggesting that the two components become decoupled after star formation has ceased. Observational studies at $z<1$ support this picture \citep{Lorenzon2025b, Spilker2025}, implying that $\delta_\mathrm{GDR}$ in quiescent galaxies may differ substantially from that of star-forming populations, likely driven by the different dust conditions (such as temperature and grain size distribution; \citealt{Cochrane2022}) and/or by the different destruction timescales \citep{Whitaker2021b}.
Determining how dust content varies with key physical properties (e.g. stellar mass, quenching timescale, environment, and AGN energetics) is essential to characterize the ISM and to distinguish among different quenching pathways. The sensitivity of ALMA2040 will enable independent, statistically robust measurements of both dust continuum emission and molecular gas in quiescent galaxies across cosmic time (down to dust masses $\mathrm{M_\mathrm{dust} \sim 10^{7.5}\, M_\odot}$ and $\mathrm{f_{dust}} = \mathrm{M_{dust}}/\mathrm{M_\star} \sim 10^{-4}$), providing a comprehensive picture of dust evolution after star formation quenching.

\subsubsection{Pinpointing molecular outflows}
\label{subsec:outflows}

Large-scale outflows driven by supermassive black holes are thought to play a central role in shutting down star formation in massive galaxies, particularly in the early Universe. In fact, at high redshift the short inferred quenching timescales require a rapid alteration of the cold gas supply or a change in its ability to form stars, potentially through ejection, heating, or phase transformation.
Although ionized-gas outflows are commonly detected in massive high-redshift galaxies, their mass content is far too low to significantly reduce or halt star formation \citep{Concas2022}. However, ionized emission lines may offer an incomplete view because they are affected by dust attenuation, particularly in the compact and highly obscured environments. Moreover, ionized gas is only one of several gas phases, and more substantial mass loss is expected in the neutral and molecular phases. JWST observations are revealing powerful neutral atomic outflows in recently quenched galaxies at $z \sim 2$–3 \citep{Belli2024, DEugenio2024}, highlighting the importance of cooler gas phases in the quenching process. Measuring the molecular gas outflow mass is now required to achieve a complete budget of the cold gas and a comprehensive understanding of the baryon cycle. 
At the moment, however, molecular outflows in quiescent galaxies are completely unconstrained; with ALMA's current capabilities, a detection would require integration times exceeding tens of days.
With ALMA2040, it will be possible to investigate the kinematics of cold gas in massive high-$z$ quiescent galaxies, probing molecular outflows down to $\mathrm{M_{H2}} \sim (2 - 5)\times 10^8\, \mathrm{M_\odot}$ (see Section~\ref{sec:technical_req}), which is comparable to the outflow mass detected in the neutral phase, clarifying their role in galaxy quenching. We highlight that a non-detection of molecular outflows would still be informative as it would indicate that AGN may quench through preventative or maintenance modes (e.g., heating the circumgalactic medium, suppressing cooling and accretion, or maintaining turbulence) rather than ejecting the molecular ISM.

\begin{figure}[t!]
    \centering
    \includegraphics[width=\linewidth]{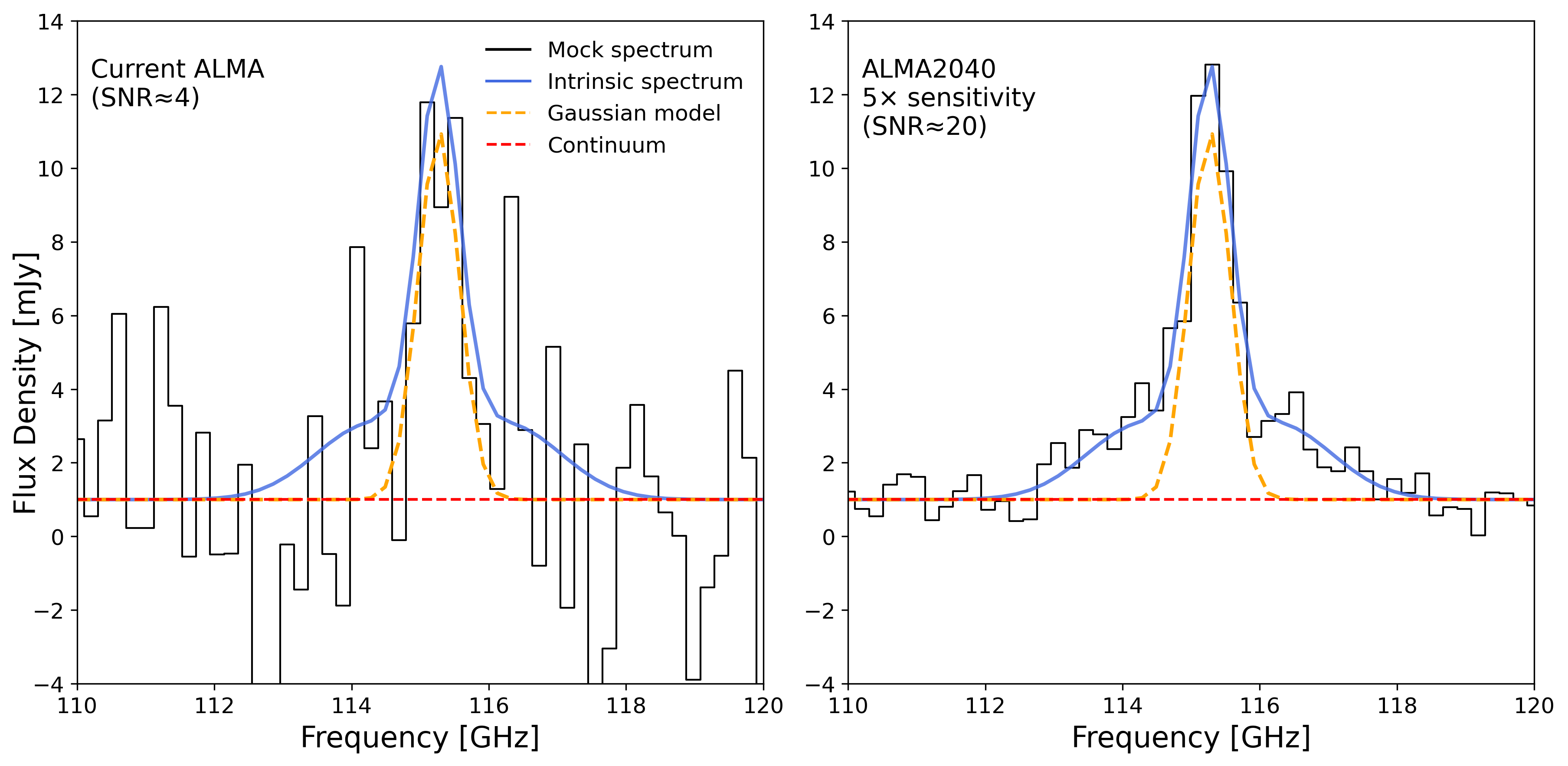}
    \caption{Mock spectrum of a galaxy showing a narrow Gaussian CO(2-1) emission and broad wings caused by an outflow along with continuum. \textbf{Left panel:} mock spectrum with integrated $\mathrm{SNR \sim 4}$, which represents the typical baseline of current ALMA observations of $z \sim 1$ quiescent galaxies with stellar mass $\mathrm{M_\star \sim 10^{10}\, M_\odot}$ and a molecular gas fraction $\mathrm{f_{gas} \gtrsim 5\%}$. \textbf{Right panel:} mock spectrum of the same galaxy as observed by ALMA2040 with a conservative sensitivity improvement of $5\times$ with respect to current ALMA.}
    \label{fig:mock_spectrum}
\end{figure}

\section{ALMA2040: Technical Requirements}
\label{sec:technical_req}

The science goals listed above require a next-generation submillimeter interferometer with the following technical requirements: a 10$\times$ increase in instantaneous spectral line sensitivity [in Band 2–9; 67–720GHz] compared to current ALMA capabilities ($\sim 8\times$ compared to anticipated ALMA-WSU capabilities) and a 10$\times$ increase in continuum sensitivity [in Band 7–10; 275–950GHz] compared to current ALMA capabilities ($\sim 5\times$ compared to anticipated ALMA-WSU capabilities). These can be achieved primarily by increasing the collecting area compared to ALMA-WSU and by achieving disruptive developments in receiver technology while at the same time providing exquisite image fidelity (through wide \textit{uv}-coverage) across a range of angular scales. Through dramatic improvements in processing speed, the correlator is no longer a limiting factor for processing the wide-(multi)bandwidth and high-spectral resolution data (same resolution as ALMA WSU) for large numbers of antennas.

Molecular gas content and outflows can be characterized by targeting CO at $z \sim 1-3$ and [CII] at $z > 3$. We aim to detect outflows with a molecular gas mass in the range $\sim (2 - 5)\times 10^8\, \mathrm{M_\odot}$ extending over $\mathrm{FWHM \sim 400 - 600\, \mathrm{km\, s^{-1}}}$, comparable to what is observed for neutral outflows in quiescent systems \citep{Belli2024, DEugenio2024} and molecular outflows in star forming galaxies \citep{Ginolfi2020, Davies2026}. Considering a [CII] luminosity-to-M$_\mathrm{H2}$ conversion factor $\alpha_\mathrm{[CII]} \sim 30\, \mathrm{M_\odot/L_\odot}$ \citep{Zanella2018, Madden2020, Dessauges-Zavadsky2020, Zhao2024, Aravena2024, Casavecchia2025}, we expect a [CII] flux in the range $0.05 - 0.01\, \mathrm{Jy\, km\, s^{-1}}$ at $z \sim 3 - 5$.
An integrated signal-to-noise ratio $\mathrm{SNR > 10}$ is needed to securely identify the outflows (Figure \ref{fig:mock_spectrum}). Therefore, we require a sensitivity in the range $\sim 0.008 - 0.0025\, \mathrm{mJy}$ over $\Delta_\mathrm{V} \sim 400 - 600\, \mathrm{km\, s^{-1}}$.
By increasing ALMA's sensitivity by $> 5\times$, molecular gas masses down to $\mathrm{M_{H2}} \sim 5\times 10^8\, \mathrm{M_\odot}$ over $\mathrm{FWHM \sim 400 - 600\, \mathrm{km\, s^{-1}}}$ at $z \sim 1 - 5$ will be probed in $2-8$h per galaxy. If the sensitivity is $10\times$ that of ALMA, the molecular gas masses that can be probed are as low as the neutral gas masses observed by JWST in the outflows from quiescent galaxies at $z\sim2$.
We need a spectral resolution of $80 - 100\, \mathrm{km\, s^{-1}}$ (at least 5 resolution elements) to probe the line width and wings.
With similar integration times, the molecular gas content of passive galaxies with stellar mass $\mathrm{M_\star \sim 10^{10} - 10^{11}\, M_\odot}$ will be observed down to gas fractions $\mathrm{f_{gas} \sim M_{H2}/M_\star} < 2\%$, i.e. $> 10\times$ below the molecular gas content of main-sequence star-forming galaxies with similar stellar masses \citep{Lin2019}. With current ALMA such observations are unfeasible as the integration time exceeds tens of days.

Furthermore, the $10\times$ increase in ALMA2040 sensitivity will allow to measure dust masses in quiescent galaxies down to $\mathrm{M_{dust} \sim 10^{7.5}\, M_\odot}$ and dust fractions down to $\mathrm{log\, f_{dust} \sim -4}$ in $< 20$ hours per galaxy. It is currently unfeasible to probe such low dust masses with current ALMA.

The main science cases (gas and dust detection, outflow characterization) require spatially integrated detections on galaxy scales.  Hence, a resolution of $\sim 0.5" - 0.8"$ achievable with intermediate or compact configuration (depending on the observing band) will suffice. However, for the brightest galaxies, a more extended configuration (resolution $\sim 0.2" - 0.5"$) will also allow the community to obtain spatially resolved measurements and constrain the morphology of the targets (Section \ref{subsec:mol_gas}).

\section{Synergies with current and future facilities}
\label{sec:synergies}
A next-generation mm/sub-mm interferometer like ALMA2040 will enable systematic measurements of both cold gas and dust in statistically representative samples of quiescent galaxies at $z \sim 1$–5, providing a complete view of the interstellar medium in galaxies closer to the primary episodes of quenching. These observations will be especially powerful in the 2040s. By then, a decade of JWST observations will have uncovered large samples of passive galaxies at high redshift and precisely measured their stellar population properties such as mass, abundances, and ages. JWST will also have systematically characterized the properties of neutral atomic gas outflows \citep{Davies2024, Taylor2026} and warm dust emission \citep{Blanquez-Sese2023, Ji2024} in large samples. Additionally, Euclid and Roman will have provided the very large-area parent samples needed to identify rare massive quiescent and transitioning systems and will have quantified their environment.  Complementary spectroscopy with ELT/HARMONI will map ionized outflows and trace the stellar kinematics (probing $\mathrm{H}\alpha$ up to $z \sim 2$ and [OIII]5007\AA\, up to $z \sim 3$). ELT/METIS will probe the warm-dust redshifted into the mid-infrared, diagnosing the buried AGN and its energetic coupling to the gas and thus demonstrating whether the black hole supplies enough energy and momentum to drive the molecular outflows detected by ALMA2040. ELT/MICADO imaging (at the ELT diffraction limit, a few milliarcseconds, with MCAO) will resolve the rest-frame optical morphology, separating merger- and interaction-driven quenching (disturbed structures, tidal features) from secular pathways. In the radio domain, SKAO and ngVLA will tightly constrain the atomic gas content in the lowest redshift quiescent targets ($z \lesssim 2$ via individual measurements and/or stacking). In particular, H\,{\sc i} and OH observations will trace atomic gas reservoirs and outflow signatures, while radio recombination lines will probe ionised gas associated with energetic feedback
processes. In favourable high-redshift or radio-bright systems, higher-frequency
SKA capabilities will also provide access to selected molecular transitions,
including CO(1-0), HCN(1-0), and HCO$^{+}$(1-0), offering complementary
constraints on cold and dense gas phases \citep{Pandey-Pommier01.2026.SKA,vanLoon01.2026.SKA}. Finally, PRIMA, if approved by NASA, will provide a complementary rest-mid-IR view of galaxies, including PAH features that, if calibrated against the molecular content, could be a valuable cross-check of the inferred molecular reservoir in quiescent galaxies. ALMA2040 will be the essential counterpart to these facilities, providing the sensitivity needed to reveal the cold dust content, molecular gas reservoirs, and molecular outflows of high-redshift quiescent galaxies. Together, these facilities will deliver a complete, multi-phase view of the gas cycle in massive quiescent galaxies, and firmly establish how quenching operates across redshift.

\section*{Acknowledgements}
This work was developed as part of the broader ALMA2040 community effort. We acknowledge the Lorentz Center and NOVA for supporting the community workshops and coordination activities that helped initiate and shape the ALMA2040 White Paper process.\\

\bibliographystyle{apj_mod}
{\bibliography{bibliography}}

@ARTICLE{Aravena2024,
       author = {{Aravena}, M. and {Heintz}, K. and {Dessauges-Zavadsky}, M. and {Oesch}, P. and {Algera}, H. and {Bouwens}, R. and {da Cunha}, E. and {Dayal}, P. and {De Looze}, I. and {Ferrara}, A. and {Fudamoto}, Y. and {Gonzalez}, V. and {Graziani}, L. and {Hygate}, A.~P.~S. and {Inami}, H. and {Pallottini}, A. and {Schneider}, R. and {Schouws}, S. and {Sommovigo}, L. and {Topping}, M. and {van der Werf}, P. and {Palla}, M.},
        title = "{The ALMA Reionization Era Bright Emission Line Survey: The molecular gas content of galaxies at z 7}",
      journal = {\aap},
         year = 2024,
        month = feb,
       volume = {682},
          eid = {A24},
        pages = {A24},
          doi = {10.1051/0004-6361/202347281},
archivePrefix = {arXiv},
       eprint = {2309.15948},
 primaryClass = {astro-ph.GA},
       adsurl = {https://ui.adsabs.harvard.edu/abs/2024A&A...682A..24A}
}

@ARTICLE{Davies2026,
       author = {{Davies}, Rebecca L. and {Fisher}, D.~B. and {Herrera-Camus}, R. and {Faisst}, A. and {Spilker}, J. and {Gonz{\'a}lez-L{\'o}pez}, J. and {Fujimoto}, S. and {Amor{\'\i}n}, R. and {Aravena}, M. and {Assef}, R.~J. and {Barcos-Mu{\~n}oz}, L. and {Boquien}, M. and {Dessauges-Zavadsky}, M. and {Ferrara}, A. and {F{\"o}rster Schreiber}, N.~M. and {Ginolfi}, M. and {G{\'o}mez-Espinoza}, D. and {Ibar}, E. and {Ikeda}, R. and {Inami}, H. and {Jones}, G.~C. and {Koekemoer}, A. and {Lee}, L.~L. and {Li}, J. and {Liu}, D. and {Liu}, Z. and {De Looze}, I. and {Mitsuhashi}, I. and {Molina}, J. and {Nanni}, A. and {Relano}, M. and {Romano}, M. and {Sawant}, P. and {Solimano}, M. and {Sommovigo}, L. and {Tadaki}, K. and {Telikova}, K. and {{\"U}bler}, H. and {Villanueva}, V. and {Wang}, W. and {Zamorani}, G.},
        title = "{Multiphase images of a powerful supernova-driven wind in the early Universe}",
      journal = {\mnras},
         year = 2026,
        month = jul,
       volume = {549},
       number = {3},
          eid = {stag874},
        pages = {stag874},
          doi = {10.1093/mnras/stag874},
archivePrefix = {arXiv},
       eprint = {2606.10271},
 primaryClass = {astro-ph.GA},
       adsurl = {https://ui.adsabs.harvard.edu/abs/2026MNRAS.549ag874D}
}

@ARTICLE{Valentino2026,
       author = {{Valentino}, F. and {Pensabene}, A. and {Weibel}, A. and {de Graaff}, A. and {Setton}, D.~J. and {Oesch}, P. and {Brammer}, G. and {Baker}, W.~M. and {Bezanson}, R. and {Greene}, J.~E. and {Heintz}, K.~E. and {Ito}, K. and {Lee}, M. and {Leja}, J. and {Matthee}, J. and {Wang}, B. and {Whitaker}, K.~E. and {Williams}, C.~C. and {Zhu}, P.},
        title = "{Extended [CII] gas emission in and around a massive quiescent galaxy at z=7.3}",
      journal = {arXiv e-prints},
         year = 2026,
        month = jun,
          eid = {arXiv:2606.21361},
        pages = {arXiv:2606.21361},
          doi = {10.48550/arXiv.2606.21361},
archivePrefix = {arXiv},
       eprint = {2606.21361},
 primaryClass = {astro-ph.GA},
       adsurl = {https://ui.adsabs.harvard.edu/abs/2026arXiv260621361V}
}

@ARTICLE{DEugenio2026,
       author = {{D'Eugenio}, C. and {Daddi}, E. and {Gobat}, R. and {Jin}, S. and {Liu}, D. and {Sun}, H. and {Gentile}, F. and {Bruckmann}, F. and {Liu}, Z. and {Delvecchio}, I. and {Vallini}, L. and {Magnelli}, B. and {Zanella}, A.},
        title = "{A first [CII] view of high-z quiescent galaxies}",
      journal = {arXiv e-prints},
         year = 2026,
        month = apr,
          eid = {arXiv:2604.09347},
        pages = {arXiv:2604.09347},
          doi = {10.48550/arXiv.2604.09347},
archivePrefix = {arXiv},
       eprint = {2604.09347},
 primaryClass = {astro-ph.GA},
       adsurl = {https://ui.adsabs.harvard.edu/abs/2026arXiv260409347D}
}

@ARTICLE{DEugenio2023,
       author = {{D'Eugenio}, C. and {Daddi}, E. and {Liu}, D. and {Gobat}, R.},
        title = "{The [CII] 158 {\ensuremath{\mu}}m emission line as a gas mass tracer in high redshift quiescent galaxies}",
      journal = {\aap},
         year = 2023,
        month = oct,
       volume = {678},
          eid = {L9},
        pages = {L9},
          doi = {10.1051/0004-6361/202347233},
archivePrefix = {arXiv},
       eprint = {2309.12730},
 primaryClass = {astro-ph.GA},
       adsurl = {https://ui.adsabs.harvard.edu/abs/2023A&A...678L...9D}
}

@ARTICLE{Umehata2025,
       author = {{Umehata}, Hideki and {Kubo}, Mariko and {Nakanishi}, Kouichiro},
        title = "{ADF22-WEB: Detection of a Molecular Gas Reservoir in a Massive Quiescent Galaxy Located in a z ≍ 3 Protocluster Core}",
      journal = {\apjl},
         year = 2025,
        month = may,
       volume = {985},
       number = {1},
          eid = {L8},
        pages = {L8},
          doi = {10.3847/2041-8213/add1d4},
archivePrefix = {arXiv},
       eprint = {2502.06538},
 primaryClass = {astro-ph.GA},
       adsurl = {https://ui.adsabs.harvard.edu/abs/2025ApJ...985L...8U}
}

@ARTICLE{French2018,
       author = {{French}, K. Decker and {Zabludoff}, Ann I. and {Yoon}, Ilsang and {Shirley}, Yancy and {Yang}, Yujin and {Smercina}, Adam and {Smith}, J.~D. and {Narayanan}, Desika},
        title = "{Why Post-starburst Galaxies Are Now Quiescent}",
      journal = {\apj},
         year = 2018,
        month = jul,
       volume = {861},
       number = {2},
          eid = {123},
        pages = {123},
          doi = {10.3847/1538-4357/aac8de},
archivePrefix = {arXiv},
       eprint = {1805.12132},
 primaryClass = {astro-ph.GA},
       adsurl = {https://ui.adsabs.harvard.edu/abs/2018ApJ...861..123F}
}

@ARTICLE{Noble2019,
       author = {{Noble}, A.~G. and {Muzzin}, A. and {McDonald}, M. and {Rudnick}, G. and {Matharu}, J. and {Cooper}, M.~C. and {Demarco}, R. and {Lidman}, C. and {Nantais}, J. and {van Kampen}, E. and {Webb}, T.~M.~A. and {Wilson}, G. and {Yee}, H.~K.~C.},
        title = "{Resolving CO (2-1) in z {\ensuremath{\sim}} 1.6 Gas-rich Cluster Galaxies with ALMA: Rotating Molecular Gas Disks with Possible Signatures of Gas Stripping}",
      journal = {\apj},
         year = 2019,
        month = jan,
       volume = {870},
       number = {2},
          eid = {56},
        pages = {56},
          doi = {10.3847/1538-4357/aaf1c6},
archivePrefix = {arXiv},
       eprint = {1809.03514},
 primaryClass = {astro-ph.GA},
       adsurl = {https://ui.adsabs.harvard.edu/abs/2019ApJ...870...56N}
}

@ARTICLE{Tacchella2015b,
       author = {{Tacchella}, S. and {Carollo}, C.~M. and {Renzini}, A. and {F{\"o}rster Schreiber}, N.~M. and {Lang}, P. and {Wuyts}, S. and {Cresci}, G. and {Dekel}, A. and {Genzel}, R. and {Lilly}, S.~J. and {Mancini}, C. and {Newman}, S. and {Onodera}, M. and {Shapley}, A. and {Tacconi}, L. and {Woo}, J. and {Zamorani}, G.},
        title = "{Evidence for mature bulges and an inside-out quenching phase 3 billion years after the Big Bang}",
      journal = {Science},
         year = 2015,
        month = apr,
       volume = {348},
       number = {6232},
        pages = {314-317},
          doi = {10.1126/science.1261094},
archivePrefix = {arXiv},
       eprint = {1504.04021},
 primaryClass = {astro-ph.GA},
       adsurl = {https://ui.adsabs.harvard.edu/abs/2015Sci...348..314T}
}

@ARTICLE{Esposito2024,
       author = {{Esposito}, Federico and {Vallini}, Livia and {Pozzi}, Francesca and {Casasola}, Viviana and {Alonso-Herrero}, Almudena and {Garc{\'\i}a-Burillo}, Santiago and {Decarli}, Roberto and {Calura}, Francesco and {Vignali}, Cristian and {Mingozzi}, Matilde and {Gruppioni}, Carlotta and {Sengupta}, Dhrubojyoti},
        title = "{Modelling molecular clouds and CO excitation in AGN-host galaxies}",
      journal = {\mnras},
         year = 2024,
        month = jan,
       volume = {527},
       number = {3},
        pages = {8727-8745},
          doi = {10.1093/mnras/stad3794},
archivePrefix = {arXiv},
       eprint = {2312.03058},
 primaryClass = {astro-ph.GA},
       adsurl = {https://ui.adsabs.harvard.edu/abs/2024MNRAS.527.8727E}
}

@ARTICLE{Vallini2019,
       author = {{Vallini}, L. and {Tielens}, A.~G.~G.~M. and {Pallottini}, A. and {Gallerani}, S. and {Gruppioni}, C. and {Carniani}, S. and {Pozzi}, F. and {Talia}, M.},
        title = "{Impact of X-rays on CO emission from high-z galaxies}",
      journal = {\mnras},
         year = 2019,
        month = dec,
       volume = {490},
       number = {4},
        pages = {4502-4514},
          doi = {10.1093/mnras/stz2837},
archivePrefix = {arXiv},
       eprint = {1910.07550},
 primaryClass = {astro-ph.GA},
       adsurl = {https://ui.adsabs.harvard.edu/abs/2019MNRAS.490.4502V}
}

@ARTICLE{Lin2019,
       author = {{Lin}, Lihwai and {Pan}, Hsi-An and {Ellison}, Sara L. and {Belfiore}, Francesco and {Shi}, Yong and {S{\'a}nchez}, Sebasti{\'a}n F. and {Hsieh}, Bau-Ching and {Rowlands}, Kate and {Ramya}, S. and {Thorp}, Mallory D. and {Li}, Cheng and {Maiolino}, Roberto},
        title = "{The ALMaQUEST Survey: The Molecular Gas Main Sequence and the Origin of the Star-forming Main Sequence}",
      journal = {\apjl},
         year = 2019,
        month = oct,
       volume = {884},
       number = {2},
          eid = {L33},
        pages = {L33},
          doi = {10.3847/2041-8213/ab4815},
archivePrefix = {arXiv},
       eprint = {1909.11243},
 primaryClass = {astro-ph.GA},
       adsurl = {https://ui.adsabs.harvard.edu/abs/2019ApJ...884L..33L}
}

@ARTICLE{Casavecchia2025,
       author = {{Casavecchia}, Benedetta and {Maio}, Umberto and {P{\'e}roux}, C{\'e}line and {Ciardi}, Benedetta},
        title = "{Atomic and molecular gas as traced by [C II] emission}",
      journal = {\aap},
         year = 2025,
        month = jan,
       volume = {693},
          eid = {A119},
        pages = {A119},
          doi = {10.1051/0004-6361/202452282},
archivePrefix = {arXiv},
       eprint = {2410.14284},
 primaryClass = {astro-ph.GA},
       adsurl = {https://ui.adsabs.harvard.edu/abs/2025A&A...693A.119C}
}

@ARTICLE{Maio2026,
       author = {{Maio}, Umberto and {P{\'e}roux}, C{\'e}line},
        title = "{The baryon budget of galaxies across the first billion years: Theoretical predictions for gas phases, depletion times, stellar return fractions}",
      journal = {\aap},
         year = 2026,
        month = apr,
       volume = {708},
          eid = {A244},
        pages = {A244},
          doi = {10.1051/0004-6361/202557624},
archivePrefix = {arXiv},
       eprint = {2603.00230},
 primaryClass = {astro-ph.GA},
       adsurl = {https://ui.adsabs.harvard.edu/abs/2026A&A...708A.244M}
}

@incollection{Pandey-Pommier01.2026.SKA, author = {Mamta Pandey-Pommier and author2 and author3 and author4 and author5},title = {Tracing gas outflows in molecular gas-rich galaxies with the SKA},year = {2026},publisher = {Advancing Astrophysics with the SKA},note = {arXiv search: Report number AASKAII/vanLoon01},booktitle = {Advancing Astrophysics with the SKA -- II (AASKAII)}}

@incollection{vanLoon01.2026.SKA, author = {Jacco van Loon and author2 and author3 and author4 and author5},title = {SKA Extragalactic spectral lines review chapter},year = {2026},publisher = {Advancing Astrophysics with the SKA},note = {arXiv search: Report number AASKAII/vanLoon01},booktitle = {Advancing Astrophysics with the SKA -- II (AASKAII)}}

@ARTICLE{Taylor2026,
       author = {{Taylor}, Elizabeth and {Carnall}, Adam C. and {Maltby}, David and {Almaini}, Omar and {Leung}, Ho-Hin and {Stevenson}, Struan D. and {Negri}, Andrea and {Cullen}, Fergus and {Wild}, Vivienne and {McLure}, Ross J. and {Shapley}, Alice E. and {Arellano-C{\'o}rdova}, Karla Z. and {Begley}, Ryan and {Bondestam}, Cecilia and {de Lisle}, Thomas and {Donnan}, Callum T. and {Dunlop}, James S. and {Ellis}, Richard and {Hewitt}, Guillaume and {Koekemoer}, Anton M. and {Liu}, Feng-Yuan Frey and {McLeod}, Derek J. and {Rowlands}, Kate and {Sanders}, Ryan L. and {Scholte}, Dirk and {Skarbinski}, Maya and {Stanton}, Thomas M.},
        title = "{The JWST EXCELS survey: outflows in 1.5 < z < 5 quiescent and recently quenched galaxies are likely relics from episodic AGN activity}",
      journal = {\mnras},
         year = 2026,
        month = jul,
       volume = {549},
       number = {3},
          eid = {stag918},
        pages = {stag918},
          doi = {10.1093/mnras/stag918},
archivePrefix = {arXiv},
       eprint = {2601.02269},
 primaryClass = {astro-ph.GA},
       adsurl = {https://ui.adsabs.harvard.edu/abs/2026MNRAS.549ag918T}
}

@ARTICLE{Davies2024,
       author = {{Davies}, Rebecca L. and {Belli}, Sirio and {Park}, Minjung and {Mendel}, J. Trevor and {Johnson}, Benjamin D. and {Conroy}, Charlie and {Benton}, Chlo{\"e} and {Bugiani}, Letizia and {Emami}, Razieh and {Leja}, Joel and {Li}, Yijia and {Maheson}, Gabriel and {Mathews}, Elijah P. and {Naidu}, Rohan P. and {Nelson}, Erica J. and {Tacchella}, Sandro and {Terrazas}, Bryan A. and {Weinberger}, Rainer},
        title = "{JWST reveals widespread AGN-driven neutral gas outflows in massive z   2 galaxies}",
      journal = {\mnras},
         year = 2024,
        month = mar,
       volume = {528},
       number = {3},
        pages = {4976-4992},
          doi = {10.1093/mnras/stae327},
archivePrefix = {arXiv},
       eprint = {2310.17939},
 primaryClass = {astro-ph.GA},
       adsurl = {https://ui.adsabs.harvard.edu/abs/2024MNRAS.528.4976D}
}

@ARTICLE{Whitaker2026,
       author = {{Whitaker}, Katherine E. and {Bezanson}, Rachel},
        title = "{Quenching of Star Formation in Massive Galaxies}",
      journal = {arXiv e-prints},
         year = 2026,
        month = jun,
          eid = {arXiv:2606.12156},
        pages = {arXiv:2606.12156},
          doi = {10.48550/arXiv.2606.12156},
archivePrefix = {arXiv},
       eprint = {2606.12156},
 primaryClass = {astro-ph.GA},
       adsurl = {https://ui.adsabs.harvard.edu/abs/2026arXiv260612156W}
}

@ARTICLE{Cochrane2022,
       author = {{Cochrane}, R.~K. and {Hayward}, C.~C. and {Angl{\'e}s-Alc{\'a}zar}, D.},
        title = "{Dust Temperature Uncertainties Hamper the Inference of Dust and Molecular Gas Masses from the Dust Continuum Emission of Quiescent High-redshift Galaxies}",
      journal = {\apjl},
         year = 2022,
        month = nov,
       volume = {939},
       number = {2},
          eid = {L27},
        pages = {L27},
          doi = {10.3847/2041-8213/ac951d},
archivePrefix = {arXiv},
       eprint = {2211.01526},
 primaryClass = {astro-ph.GA},
       adsurl = {https://ui.adsabs.harvard.edu/abs/2022ApJ...939L..27C}
}

@ARTICLE{Zhao2024,
       author = {{Zhao}, Yinghe and {Liu}, Jiamin and {Zhang}, Zhi-Yu and {Bisbas}, Thomas G.},
        title = "{Ionized Carbon in Galaxies: The [C II] 158 {\ensuremath{\mu}}m Line as a Total Molecular Gas Mass Tracer Revisited}",
      journal = {\apj},
         year = 2024,
        month = dec,
       volume = {977},
       number = {1},
          eid = {46},
        pages = {46},
          doi = {10.3847/1538-4357/ad8c3e},
archivePrefix = {arXiv},
       eprint = {2410.20684},
 primaryClass = {astro-ph.GA},
       adsurl = {https://ui.adsabs.harvard.edu/abs/2024ApJ...977...46Z}
}

@ARTICLE{Xu2026,
       author = {{Xu}, Ke and {Wang}, Tao and {Daddi}, Emanuele and {Elbaz}, David and {Sun}, Hanwen and {Chen}, Longyue and {Hao}, Qiaoyang and {Gobat}, Raphael and {Zanella}, Anita and {Liu}, Daizhong and {Xiao}, Mengyuan and {Cen}, Renyue and {Kodama}, Tadayuki and {Kohno}, Kotaro and {Yang}, Tiancheng and {Xu}, Can and {Zhang}, Zhi-Yu and {Zhou}, Luwenjia and {Valentino}, Francesco},
        title = "{Ram-pressure stripping caught in action in a young cluster at z = 2.51}",
      journal = {Nature Astronomy},
         year = 2026,
        month = jun,
          doi = {10.1038/s41550-026-02892-9},
       adsurl = {https://ui.adsabs.harvard.edu/abs/2026NatAs.tmp..134X}
}

@ARTICLE{Donofrio2026,
       author = {{D'Onofrio}, Vincenzo R. and {Spilker}, Justin S. and {Bezanson}, Rachel and {Feldmann}, Robert and {Goulding}, Andy D. and {Greene}, Jenny E. and {Kriek}, Mariska and {Kumar}, Anika and {Luo}, Yuanze and {Narayanan}, Desika and {Setton}, David J. and {Suess}, Katherine A. and {Verrico}, Margaret E.},
        title = "{Molecular Gas Excitation in z {\ensuremath{\sim}} 0.7 Gas-rich Poststarburst Galaxies from SQuIGGL{\textrightarrow}E}",
      journal = {\apj},
         year = 2026,
        month = may,
       volume = {1003},
       number = {1},
          eid = {52},
        pages = {52},
          doi = {10.3847/1538-4357/ae64e7},
archivePrefix = {arXiv},
       eprint = {2602.17766},
 primaryClass = {astro-ph.GA},
       adsurl = {https://ui.adsabs.harvard.edu/abs/2026ApJ..1003...52D}
}

@ARTICLE{Zanella2026,
       author = {{Zanella}, A. and {Belli}, S. and {Valentino}, F.~M. and {Bolamperti}, A.},
        title = "{Gas excitation of post-SB galaxies at 0.6 < z < 1.3}",
      journal = {\aap},
         year = 2026,
        month = apr,
       volume = {709},
          eid = {A14},
        pages = {A14},
          doi = {10.1051/0004-6361/202558073},
archivePrefix = {arXiv},
       eprint = {2511.08829},
 primaryClass = {astro-ph.GA},
       adsurl = {https://ui.adsabs.harvard.edu/abs/2026A&A...709A..14Z}
}

@ARTICLE{vandeVoort2018,
       author = {{van de Voort}, Freeke and {Davis}, Timothy A. and {Matsushita}, Satoki and {Rowlands}, Kate and {Shabala}, Stanislav S. and {Allison}, James R. and {Ting}, Yuan-Sen and {Sansom}, Anne E. and {van der Werf}, Paul P.},
        title = "{An ALMA view of star formation efficiency suppression in early-type galaxies after gas-rich minor mergers}",
      journal = {\mnras},
         year = 2018,
        month = may,
       volume = {476},
       number = {1},
        pages = {122-132},
          doi = {10.1093/mnras/sty228},
archivePrefix = {arXiv},
       eprint = {1801.08140},
 primaryClass = {astro-ph.GA},
       adsurl = {https://ui.adsabs.harvard.edu/abs/2018MNRAS.476..122V}
}

@ARTICLE{Ji2024,
       author = {{Ji}, Zhiyuan and {Williams}, Christina C. and {Rieke}, George H. and {Lyu}, Jianwei and {Alberts}, Stacey and {Sun}, Fengwu and {Helton}, Jakob M. and {Rieke}, Marcia and {Shivaei}, Irene and {D'Eugenio}, Francesco and {Tacchella}, Sandro and {Robertson}, Brant and {Zhu}, Yongda and {Maiolino}, Roberto and {Bunker}, Andrew J. and {Sun}, Yang and {Willmer}, Christopher N.~A.},
        title = "{Extended hot dust emission around the earliest massive quiescent galaxy}",
      journal = {arXiv e-prints},
         year = 2024,
        month = sep,
          eid = {arXiv:2409.17233},
        pages = {arXiv:2409.17233},
          doi = {10.48550/arXiv.2409.17233},
archivePrefix = {arXiv},
       eprint = {2409.17233},
 primaryClass = {astro-ph.GA},
       adsurl = {https://ui.adsabs.harvard.edu/abs/2024arXiv240917233J}
}

@ARTICLE{deGraaff2025,
       author = {{de Graaff}, Anna and {Setton}, David J. and {Brammer}, Gabriel and {Cutler}, Sam and {Suess}, Katherine A. and {Labb{\'e}}, Ivo and {Leja}, Joel and {Weibel}, Andrea and {Maseda}, Michael V. and {Whitaker}, Katherine E. and {Bezanson}, Rachel and {Boogaard}, Leindert A. and {Cleri}, Nikko J. and {De Lucia}, Gabriella and {Franx}, Marijn and {Greene}, Jenny E. and {Hirschmann}, Michaela and {Matthee}, Jorryt and {McConachie}, Ian and {Naidu}, Rohan P. and {Oesch}, Pascal A. and {Price}, Sedona H. and {Rix}, Hans-Walter and {Valentino}, Francesco and {Wang}, Bingjie and {Williams}, Christina C.},
        title = "{Efficient formation of a massive quiescent galaxy at redshift 4.9}",
      journal = {Nat. As.},
         year = 2025,
        month = feb,
       volume = {9},
        pages = {280-292},
          doi = {10.1038/s41550-024-02424-3},
archivePrefix = {arXiv},
       eprint = {2404.05683},
 primaryClass = {astro-ph.GA},
       adsurl = {https://ui.adsabs.harvard.edu/abs/2025NatAs...9..280D}
}

@ARTICLE{Weibel2025,
       author = {{Weibel}, Andrea and {de Graaff}, Anna and {Setton}, David J. and {Miller}, Tim B. and {Oesch}, Pascal A. and {Brammer}, Gabriel and {Lagos}, Claudia D.~P. and {Whitaker}, Katherine E. and {Williams}, Christina C. and {Baggen}, Josephine F.~W. and {Bezanson}, Rachel and {Boogaard}, Leindert A. and {Cleri}, Nikko J. and {Greene}, Jenny E. and {Hirschmann}, Michaela and {Hviding}, Raphael E. and {Kuruvanthodi}, Adarsh and {Labb{\'e}}, Ivo and {Leja}, Joel and {Maseda}, Michael V. and {Matthee}, Jorryt and {McConachie}, Ian and {Naidu}, Rohan P. and {Roberts-Borsani}, Guido and {Schaerer}, Daniel and {Suess}, Katherine A. and {Valentino}, Francesco and {van Dokkum}, Pieter and {Wang}, Bingjie},
        title = "{RUBIES Reveals a Massive Quiescent Galaxy at z = 7.3}",
      journal = {\apj},
         year = 2025,
        month = apr,
       volume = {983},
       number = {1},
          eid = {11},
        pages = {11},
          doi = {10.3847/1538-4357/adab7a},
archivePrefix = {arXiv},
       eprint = {2409.03829},
 primaryClass = {astro-ph.GA},
       adsurl = {https://ui.adsabs.harvard.edu/abs/2025ApJ...983...11W}
}

@ARTICLE{Scholtz2024,
       author = {{Scholtz}, Jan and {D'Eugenio}, Francesco and {Maiolino}, Roberto and {P{\'e}rez-Gonz{\'a}lez}, Pablo G. and {Circosta}, Chiara and {Tacchella}, Sandro and {Williams}, Christina C. and {Alberts}, Stacey and {Arribas}, Santiago and {Baker}, William M. and {Bertola}, Elena and {Bunker}, Andrew J. and {Carniani}, Stefano and {Charlot}, Stephane and {Cresci}, Giovanni and {Jones}, Gareth C. and {Kumari}, Nimisha and {Lamperti}, Isabella and {Looser}, Tobias J. and {Pino}, Bruno Rodr{\'\i}guez Del and {Robertson}, Brant and {Parlanti}, Eleonora and {Perna}, Michele and {{\"U}bler}, Hannah and {Venturi}, Giacomo and {Witstok}, Joris},
        title = "{Measurement of the gas consumption history of a massive quiescent galaxy}",
      journal = {Nature Astronomy},
         year = 2026,
        month = mar,
       volume = {10},
        pages = {431-439},
          doi = {10.1038/s41550-025-02751-z},
archivePrefix = {arXiv},
       eprint = {2405.19401},
 primaryClass = {astro-ph.GA},
       adsurl = {https://ui.adsabs.harvard.edu/abs/2026NatAs..10..431S}
}

@ARTICLE{Lee2024,
       author = {{Lee}, Minju M. and {Steidel}, Charles C. and {Brammer}, Gabriel and {F{\"o}rster-Schreiber}, Natascha and {Renzini}, Alvio and {Liu}, Daizhong and {Herrera-Camus}, Rodrigo and {Naab}, Thorsten and {Price}, Sedona H. and {{\"U}bler}, Hannah and {Arriagada-Neira}, Sebasti{\'a}n and {Magdis}, Georgios},
        title = "{High dust content of a quiescent galaxy at z   2 revealed by deep ALMA observation}",
      journal = {\mnras},
         year = 2024,
        month = feb,
       volume = {527},
       number = {4},
        pages = {9529-9547},
          doi = {10.1093/mnras/stad3718},
archivePrefix = {arXiv},
       eprint = {2311.00023},
 primaryClass = {astro-ph.GA},
       adsurl = {https://ui.adsabs.harvard.edu/abs/2024MNRAS.527.9529L}
}

@ARTICLE{Whitaker2021a,
       author = {{Whitaker}, Katherine E. and {Williams}, Christina C. and {Mowla}, Lamiya and {Spilker}, Justin S. and {Toft}, Sune and {Narayanan}, Desika and {Pope}, Alexandra and {Magdis}, Georgios E. and {van Dokkum}, Pieter G. and {Akhshik}, Mohammad and {Bezanson}, Rachel and {Brammer}, Gabriel B. and {Leja}, Joel and {Man}, Allison and {Nelson}, Erica J. and {Richard}, Johan and {Pacifici}, Camilla and {Sharon}, Keren and {Valentino}, Francesco},
        title = "{Quenching of star formation from a lack of inflowing gas to galaxies}",
      journal = {Nat.},
         year = 2021,
        month = sep,
       volume = {597},
       number = {7877},
        pages = {485-488},
          doi = {10.1038/s41586-021-03806-7},
archivePrefix = {arXiv},
       eprint = {2109.10384},
 primaryClass = {astro-ph.GA},
       adsurl = {https://ui.adsabs.harvard.edu/abs/2021Natur.597..485W}
}

@ARTICLE{Whitaker2021b,
       author = {{Whitaker}, Katherine E. and {Narayanan}, Desika and {Williams}, Christina C. and {Li}, Qi and {Spilker}, Justin S. and {Dav{\'e}}, Romeel and {Akhshik}, Mohammad and {Akins}, Hollis B. and {Bezanson}, Rachel and {Katz}, Neal and {Leja}, Joel and {Magdis}, Georgios E. and {Mowla}, Lamiya and {Nelson}, Erica J. and {Pope}, Alexandra and {Privon}, George C. and {Toft}, Sune and {Valentino}, Francesco},
        title = "{High Molecular-gas to Dust Mass Ratios Predicted in Most Quiescent Galaxies}",
      journal = {\apjl},
         year = 2021,
        month = dec,
       volume = {922},
       number = {2},
          eid = {L30},
        pages = {L30},
          doi = {10.3847/2041-8213/ac399f},
archivePrefix = {arXiv},
       eprint = {2111.05349},
 primaryClass = {astro-ph.GA},
       adsurl = {https://ui.adsabs.harvard.edu/abs/2021ApJ...922L..30W}
}

@ARTICLE{Blanquez-Sese2023,
       author = {{Bl{\'a}nquez-Ses{\'e}}, D. and {G{\'o}mez-Guijarro}, C. and {Magdis}, G.~E. and {Magnelli}, B. and {Gobat}, R. and {Daddi}, E. and {Franco}, M. and {Whitaker}, K. and {Valentino}, F. and {Adscheid}, S. and {Schinnerer}, E. and {Zanella}, A. and {Xiao}, M. and {Wang}, T. and {Liu}, D. and {Kokorev}, V. and {Elbaz}, D.},
        title = "{The gas mass reservoir of quiescent galaxies at cosmic noon}",
      journal = {\aap},
         year = 2023,
        month = jun,
       volume = {674},
          eid = {A166},
        pages = {A166},
          doi = {10.1051/0004-6361/202345977},
archivePrefix = {arXiv},
       eprint = {2303.12110},
 primaryClass = {astro-ph.GA},
       adsurl = {https://ui.adsabs.harvard.edu/abs/2023A&A...674A.166B}
}

@ARTICLE{Concas2022,
       author = {{Concas}, Alice and {Maiolino}, Roberto and {Curti}, Mirko and {Hayden-Pawson}, Connor and {Cirasuolo}, Michele and {Jones}, Gareth C. and {Mercurio}, Amata and {Belfiore}, Francesco and {Cresci}, Giovanni and {Cullen}, Fergus and {Mannucci}, Filippo and {Marconi}, Alessandro and {Cappellari}, Michele and {Cicone}, Claudia and {Peng}, Yingjie and {Troncoso}, Paulina},
        title = "{Being KLEVER at cosmic noon: Ionized gas outflows are inconspicuous in low-mass star-forming galaxies but prominent in massive AGN hosts}",
      journal = {\mnras},
         year = 2022,
        month = jun,
       volume = {513},
       number = {2},
        pages = {2535-2562},
          doi = {10.1093/mnras/stac1026},
archivePrefix = {arXiv},
       eprint = {2203.11958},
 primaryClass = {astro-ph.GA},
       adsurl = {https://ui.adsabs.harvard.edu/abs/2022MNRAS.513.2535C}
}

@ARTICLE{Lorenzon2025b,
       author = {{Lorenzon}, Giuliano and {Donevski}, Darko and {Man}, Allison W.~S. and {Romano}, Michael and {Whitaker}, Katherine E. and {Belli}, Sirio and {Liu}, Daizhong and {Lee}, Minju M. and {Narayanan}, Desika and {Long}, Arianna and {Shivaei}, Irene and {Nanni}, Ambra and {Lisiecki}, Krzysztof and {Sawant}, Prasad and {Rodighiero}, Giulia and {Damjanov}, Ivana and {Junais} and {Dave}, Romeel and {Pappalardo}, Ciro and {Lovell}, Christopher and {Hamed}, Mahmoud},
        title = "{ALMA reveals diverse dust-to-gas mass ratios and quenching modes in old quiescent galaxies}",
      journal = {arXiv e-prints},
         year = 2025,
        month = sep,
          eid = {arXiv:2509.10079},
        pages = {arXiv:2509.10079},
          doi = {10.48550/arXiv.2509.10079},
archivePrefix = {arXiv},
       eprint = {2509.10079},
 primaryClass = {astro-ph.GA},
       adsurl = {https://ui.adsabs.harvard.edu/abs/2025arXiv250910079L}
}

@ARTICLE{Lorenzon2025a,
       author = {{Lorenzon}, G. and {Donevski}, D. and {Lisiecki}, K. and {Lovell}, C. and {Romano}, M. and {Narayanan}, D. and {Dav{\'e}}, R. and {Man}, A. and {Whitaker}, K.~E. and {Nanni}, A. and {Long}, A. and {Lee}, M.~M. and {Junais} and {Ma{\l}ek}, K. and {Rodighiero}, G. and {Li}, Q.},
        title = "{Tracing the evolutionary pathways of dust and cold gas in high-z quiescent galaxies with SIMBA}",
      journal = {\aap},
         year = 2025,
        month = jan,
       volume = {693},
          eid = {A118},
        pages = {A118},
          doi = {10.1051/0004-6361/202450393},
archivePrefix = {arXiv},
       eprint = {2404.10568},
 primaryClass = {astro-ph.GA},
       adsurl = {https://ui.adsabs.harvard.edu/abs/2025A&A...693A.118L}
}

@ARTICLE{Spilker2025,
       author = {{Spilker}, Justin S. and {Whitaker}, Katherine E. and {Narayanan}, Desika and {Bezanson}, Rachel and {Bodansky}, Sarah and {D'Onofrio}, Vincenzo R. and {Feldmann}, Robert and {Goulding}, Andy D. and {Greene}, Jenny E. and {Kriek}, Mariska and {Luo}, Yuanze and {Setton}, David J. and {Suess}, Katherine A. and {van der Wel}, Arjen and {Verrico}, Margaret E. and {Williams}, Christina C. and {Woodrum}, Charity and {Wu}, Po-Feng},
        title = "{Unusually High Gas-to-dust Ratios Observed in High-redshift Quiescent Galaxies}",
      journal = {\apjl},
         year = 2025,
        month = nov,
       volume = {993},
       number = {2},
          eid = {L40},
        pages = {L40},
          doi = {10.3847/2041-8213/ae14d6},
       adsurl = {https://ui.adsabs.harvard.edu/abs/2025ApJ...993L..40S}
}

@ARTICLE{Setton2025,
       author = {{Setton}, David J. and {Spilker}, Justin S. and {Bezanson}, Rachel and {Suess}, Katherine A. and {Greene}, Jenny E. and {Goulding}, Andy D. and {Cenci}, Elia and {D'Onofrio}, Vincenzo R. and {Feldmann}, Robert and {Kriek}, Mariska and {Kumar}, Anika and {Luo}, Yuanze and {Narayanan}, Desika and {Verrico}, Margaret E. and {Zhu}, Pengpei},
        title = "{SQuIGG$\vec{L}$E: Buried star formation cannot explain the rapidly fading CO(2-1) luminosity in massive, $z\sim0.7$ post-starburst galaxies}",
      journal = {arXiv e-prints},
         year = 2025,
        month = aug,
          eid = {arXiv:2509.00148},
        pages = {arXiv:2509.00148},
          doi = {10.48550/arXiv.2509.00148},
archivePrefix = {arXiv},
       eprint = {2509.00148},
 primaryClass = {astro-ph.GA},
       adsurl = {https://ui.adsabs.harvard.edu/abs/2025arXiv250900148S}
}

@ARTICLE{DEugenio2024,
       author = {{D'Eugenio}, Francesco and {P{\'e}rez-Gonz{\'a}lez}, Pablo G. and {Maiolino}, Roberto and {Scholtz}, Jan and {Perna}, Michele and {Circosta}, Chiara and {{\"U}bler}, Hannah and {Arribas}, Santiago and {B{\"o}ker}, Torsten and {Bunker}, Andrew J. and {Carniani}, Stefano and {Charlot}, Stephane and {Chevallard}, Jacopo and {Cresci}, Giovanni and {Curtis-Lake}, Emma and {Jones}, Gareth C. and {Kumari}, Nimisha and {Lamperti}, Isabella and {Looser}, Tobias J. and {Parlanti}, Eleonora and {Rix}, Hans-Walter and {Robertson}, Brant and {Rodr{\'\i}guez Del Pino}, Bruno and {Tacchella}, Sandro and {Venturi}, Giacomo and {Willott}, Chris J.},
        title = "{A fast-rotator post-starburst galaxy quenched by supermassive black-hole feedback at z = 3}",
      journal = {Nat. As.},
         year = 2024,
        month = nov,
       volume = {8},
        pages = {1443-1456},
          doi = {10.1038/s41550-024-02345-1},
archivePrefix = {arXiv},
       eprint = {2308.06317},
 primaryClass = {astro-ph.GA},
       adsurl = {https://ui.adsabs.harvard.edu/abs/2024NatAs...8.1443D}
}

@ARTICLE{Belli2024,
       author = {{Belli}, Sirio and {Park}, Minjung and {Davies}, Rebecca L. and {Mendel}, J. Trevor and {Johnson}, Benjamin D. and {Conroy}, Charlie and {Benton}, Chlo{\"e} and {Bugiani}, Letizia and {Emami}, Razieh and {Leja}, Joel and {Li}, Yijia and {Maheson}, Gabriel and {Mathews}, Elijah P. and {Naidu}, Rohan P. and {Nelson}, Erica J. and {Tacchella}, Sandro and {Terrazas}, Bryan A. and {Weinberger}, Rainer},
        title = "{Star formation shut down by multiphase gas outflow in a galaxy at a redshift of 2.45}",
      journal = {Nat.},
         year = 2024,
        month = jun,
       volume = {630},
       number = {8015},
        pages = {54-58},
          doi = {10.1038/s41586-024-07412-1},
archivePrefix = {arXiv},
       eprint = {2308.05795},
 primaryClass = {astro-ph.GA},
       adsurl = {https://ui.adsabs.harvard.edu/abs/2024Natur.630...54B}
}

@ARTICLE{Carnall2023,
       author = {{Carnall}, Adam C. and {McLure}, Ross J. and {Dunlop}, James S. and {McLeod}, Derek J. and {Wild}, Vivienne and {Cullen}, Fergus and {Magee}, Dan and {Begley}, Ryan and {Cimatti}, Andrea and {Donnan}, Callum T. and {Hamadouche}, Massissilia L. and {Jewell}, Sophie M. and {Walker}, Sam},
        title = "{A massive quiescent galaxy at redshift 4.658}",
      journal = {Nat.},
         year = 2023,
        month = jul,
       volume = {619},
       number = {7971},
        pages = {716-719},
          doi = {10.1038/s41586-023-06158-6},
archivePrefix = {arXiv},
       eprint = {2301.11413},
 primaryClass = {astro-ph.GA},
       adsurl = {https://ui.adsabs.harvard.edu/abs/2023Natur.619..716C}
}

@ARTICLE{Zanella2023,
       author = {{Zanella}, A. and {Valentino}, F. and {Gallazzi}, A. and {Belli}, S. and {Magdis}, G. and {Bolamperti}, A.},
        title = "{The large molecular gas fraction of post-starburst galaxies at $z > 1$}",
      journal = {\mnras},
         year = 2023,
        month = sep,
       volume = {524},
       number = {1},
        pages = {923-939},
          doi = {10.1093/mnras/stad1821},
archivePrefix = {arXiv},
       eprint = {2306.08120},
 primaryClass = {astro-ph.GA},
       adsurl = {https://ui.adsabs.harvard.edu/abs/2023MNRAS.524..923Z}
}

@ARTICLE{Suess2017,
       author = {{Suess}, Katherine A. and {Bezanson}, Rachel and {Spilker}, Justin S. and {Kriek}, Mariska and {Greene}, Jenny E. and {Feldmann}, Robert and {Hunt}, Qiana and {Narayanan}, Desika},
        title = "{Massive Quenched Galaxies at z {\ensuremath{\sim}} 0.7 Retain Large Molecular Gas Reservoirs}",
      journal = {\apjl},
         year = 2017,
        month = sep,
       volume = {846},
       number = {2},
          eid = {L14},
        pages = {L14},
          doi = {10.3847/2041-8213/aa85dc},
archivePrefix = {arXiv},
       eprint = {1708.03337},
 primaryClass = {astro-ph.GA},
       adsurl = {https://ui.adsabs.harvard.edu/abs/2017ApJ...846L..14S}
}

@ARTICLE{Man2018,
       author = {{Man}, Allison and {Belli}, Sirio},
        title = "{Star formation quenching in massive galaxies}",
      journal = {Nat. As.},
         year = 2018,
        month = sep,
       volume = {2},
        pages = {695-697},
          doi = {10.1038/s41550-018-0558-1},
archivePrefix = {arXiv},
       eprint = {1809.00722},
 primaryClass = {astro-ph.GA},
       adsurl = {https://ui.adsabs.harvard.edu/abs/2018NatAs...2..695M}
}

@ARTICLE{Madden2020,
       author = {{Madden}, S.~C. and {Cormier}, D. and {Hony}, S. and {Lebouteiller}, V. and {Abel}, N. and {Galametz}, M. and {De Looze}, I. and {Chevance}, M. and {Polles}, F.~L. and {Lee}, M. -Y. and {Galliano}, F. and {Lambert-Huyghe}, A. and {Hu}, D. and {Ramambason}, L.},
        title = "{Tracing the total molecular gas in galaxies: [CII] and the CO-dark gas}",
      journal = {\aap},
         year = 2020,
        month = nov,
       volume = {643},
          eid = {A141},
        pages = {A141},
          doi = {10.1051/0004-6361/202038860},
archivePrefix = {arXiv},
       eprint = {2009.00649},
 primaryClass = {astro-ph.GA},
       adsurl = {https://ui.adsabs.harvard.edu/abs/2020A&A...643A.141M}
}

@ARTICLE{Dessauges-Zavadsky2020,
       author = {{Dessauges-Zavadsky}, M. and {Ginolfi}, M. and {Pozzi}, F. and {B{\'e}thermin}, M. and {Le F{\`e}vre}, O. and {Fujimoto}, S. and {Silverman}, J.~D. and {Jones}, G.~C. and {Vallini}, L. and {Schaerer}, D. and {Faisst}, A.~L. and {Khusanova}, Y. and {Fudamoto}, Y. and {Cassata}, P. and {Loiacono}, F. and {Capak}, P.~L. and {Yan}, L. and {Amorin}, R. and {Bardelli}, S. and {Boquien}, M. and {Cimatti}, A. and {Gruppioni}, C. and {Hathi}, N.~P. and {Ibar}, E. and {Koekemoer}, A.~M. and {Lemaux}, B.~C. and {Narayanan}, D. and {Oesch}, P.~A. and {Rodighiero}, G. and {Romano}, M. and {Talia}, M. and {Toft}, S. and {Vergani}, D. and {Zamorani}, G. and {Zucca}, E.},
        title = "{The ALPINE-ALMA [C II] survey. Molecular gas budget in the early Universe as traced by [C II]}",
      journal = {\aap},
         year = 2020,
        month = nov,
       volume = {643},
          eid = {A5},
        pages = {A5},
          doi = {10.1051/0004-6361/202038231},
archivePrefix = {arXiv},
       eprint = {2004.10771},
 primaryClass = {astro-ph.GA},
       adsurl = {https://ui.adsabs.harvard.edu/abs/2020A&A...643A...5D}
}

@ARTICLE{Gobat2018,
       author = {{Gobat}, R. and {Daddi}, E. and {Magdis}, G. and {Bournaud}, F. and {Sargent}, M. and {Martig}, M. and {Jin}, S. and {Finoguenov}, A. and {B{\'e}thermin}, M. and {Hwang}, H.~S. and {Renzini}, A. and {Wilson}, G.~W. and {Aretxaga}, I. and {Yun}, M. and {Strazzullo}, V. and {Valentino}, F.},
        title = "{The unexpectedly large dust and gas content of quiescent galaxies at z > 1.4}",
      journal = {Nat. As.},
         year = 2018,
        month = jan,
       volume = {2},
        pages = {239-246},
          doi = {10.1038/s41550-017-0352-5},
archivePrefix = {arXiv},
       eprint = {1703.02207},
 primaryClass = {astro-ph.GA},
       adsurl = {https://ui.adsabs.harvard.edu/abs/2018NatAs...2..239G}
}

@ARTICLE{Papadopoulos2004,
   author = {{Papadopoulos}, P.~P. and {Greve}, T.~R.},
    title = "{C I Emission in Ultraluminous Infrared Galaxies as a Molecular Gas Mass Tracer}",
  journal = {\apjl},
   eprint = {astro-ph/0409559},
     year = 2004,
    month = nov,
   volume = 615,
    pages = {L29-L32},
      doi = {10.1086/426059},
   adsurl = {http://adsabs.harvard.edu/abs/2004ApJ...615L..29P}
}

\vspace{1em}
{
\footnotesize{
\noindent $^1$ Istituto Nazionale di Astrofisica (INAF), Via Gobetti 93/3, 40129, Bologna, Italy \\
$^2$ Dipartimento di Fisica e Astronomia, Università di Bologna, Bologna, Italy \\
$^3$ SKA Observatory, Jodrell Bank, SK11 9FT, UK \\
$^4$ Instituto de Estudios Astrofísicos, Facultad de Ingeniería y Ciencias, Universidad Diego Portales, Avenida Ejército 441, Santiago, Chile \\
$^5$ Millenium Nucleus for Galaxies (MINGAL), Chile \\
$^6$ UK ALMA Regional Centre Node, Jodrell Bank Centre for Astrophysics, Department of Physics and Astronomy, The University of Manchester, Oxford Road, Manchester M13 9PL, UK \\
$^7$ Department of Physics and Astronomy and PITT PACC, University of Pittsburgh, Pittsburgh, PA 15260, USA \\
$^8$ Scuola Normale Superiore, Piazza dei Cavalieri 7, I-56126 Pisa, Italy \\
$^9$ Observatoire de Paris, LUX, Collège de France, CNRS, PSL University, Sorbonne University, 75014, Paris, France \\
$^{10}$ Sorbonne Universite\`e, UPMC Paris 6 and CNRS, UMR 7095, Institut d’Astrophysique de Paris, 98b bd. Arago, 75014 Paris, France \\
$^{11}$ Max-Planck-Institut f\"ur Astronomie, K\"onigstuhl 17, D-69117 Heidelberg, Germany \\
$^{12}$ Dpto. de F\'isica Te\'orica y del Cosmos, Edificio Mecenas, Campus de Fuentenueva, Universidad de Granada, E-18071, Granada, Spain  \\
$^{13}$ Instituto Carlos I de F\'isica Te\'orica y Computacional, Universidad de Granada, E-18071, Granada, Spain \\
$^{14}$ IPAC, California Institute of Technology, 1200 E. California Blvd. Pasadena, CA 91125, USA \\
$^{15}$ Department of Astronomy, University of Illinois, 1002 W. Green St., Urbana, IL 61801, USA \\ 
$^{16}$ Center for Astrophysical Surveys, National Center for Supercomputing Applications, Urbana, IL 61801, USA \\
$^{17}$ Observatoire astronomique, Université de Strasbourg, CNRS UMR 7550, Strasbourg, France \\
$^{18}$ Centro de Astrobiología (CAB), CSIC-INTA, Ctra. de Ajalvir km 4, Torrejón de Ardoz, E-28850, Madrid, Spain \\
$^{19}$ Steward Observatory, University of Arizona, 933 N. Cherry Avenue, Tucson, AZ 85721, USA \\
$^{20}$ Kavli Institute for Cosmology, University of Cambridge, Madingley Road, Cambridge CB3 0HA, UK \\
$^{21}$ Cavendish Laboratory, University of Cambridge, 19 JJ Thomson Avenue, Cambridge CB3 0HE, UK \\
$^{22}$ Centre for Astrophysics \& Supercomputing, Swinburne University of Technology, Hawthorn, VIC 3122, Australia \\ 
$^{23}$ ARC Centre of Excellence for Dark Matter Particle Physics, Australia \\
$^{24}$ Instituto de Astronomía y Ciencias Planetarias, Universidad de Atacama, Copayapu 485, Copiapó, Chile \\
$^{25}$ SISSA, Via Bonomea 265, Trieste, Italy \\
$^{26}$ Centro de Astrobiología (CAB), CSIC-INTA, Carretera de Ajalvir km 4, Torrejón de Ardoz, 28850, Madrid, Spain \\
$^{27}$ Italian National Institute of Astrophysics - Observatory of Trieste, via G.~Tieplo 11, 34141 Trieste (Italy) \\
$^{28}$ Istituto Nazionale di Astrofisica (INAF), Osservatorio di Padova, vicolo dell'Osservatorio 5, 35122 Padova (Italy) \\
$^{29}$ Department of Astronomy, University of Michigan, 1085 South University Avenue, Ann Arbor, MI 48109, USA \\
$^{30}$ Pole Scientific, University Catholic of Lyon- University of Lyon, 10 place des Archives 69002, Lyon, France \\
$^{31}$ European Southern Observatory, Karl-Schwarzschild-Str. 2, 85748, Garching, Germany \\
$^{32}$ Institut f\"ur Astrophysik, Universit\"at zu K\"oln, Germany \\
$^{33}$ Dept. of Molecular Astrophysics. IFF-CSIC; C/ Serrano 123, 28006, Madrid, Spain \\
$^{34}$ Jet Propulsion Laboratory, California Institute of Technology, 4800 Oak Grove Drive, Pasadena, CA 91109, US \\
$^{35}$ Department of Physics and Astronomy and George P. and Cynthia Woods Mitchell Institute for Fundamental Physics and Astronomy, Texas A\&M University, 4242 TAMU, College Station, TX 77843-4242, USA \\
$^{36}$ Department of Astrophysics, University of Vienna, Türkenschanzstraße 17, 1180, Vienna, Austria \\
$^{37}$ Cosmic Dawn Center (DAWN), Denmark \\
$^{38}$ DTU Space, Technical University of Denmark, Elektrovej 327, DK-2800, Kgs. Lyngby, Denmark \\
$^{39}$ Lennard-Jones Laboratories, Keele University, ST5 5BG, UK \\
$^{40}$ NSF National Optical-Infrared Astronomy Research Laboratory, 950 North Cherry Avenue, Tucson, AZ 85719, USA
}
}
\end{document}